\documentclass{article}

\usepackage{amsmath,amssymb,amsthm,mathtools,listings,multirow,bbm}%,dirtree,boxedminipage,verbatim,anyfontsize,balance,amsthm,subcaption,epstopdf,cprotect,astyped,color,booktabs,graphicx,stmaryrd,
\usepackage[all]{xy}
\newlength{\mysidemargin}
\newtheorem{theorem}{Theorem}

\def\mytitle{Owicki--Gries Logic for Timestamp~Semantics}

\def\myabstract{

  Whereas an extension with non-interference of Hoare logic for
  sequential programs Owicki--Gries logic ensures the correctness of
  concurrent programs on strict consistency, it is unsound to weak
  memory models adopted by modern computer architectures and
  specifications of programming languages. This paper proposes a novel
  non-interference notion and provides concurrent program logic sound
  to timestamp semantics corresponding to a weak memory model that
  allows delays in the effects of store instructions. This paper
  reports three theoretically interesting techniques for modifying
  non-interference to support delays in the effects of store
  instructions. The techniques contribute to a better understanding of
  constructing concurrent program logic.

}
\def\mykeywords{program logic, weak memory model, non-interference, timestamp, vector clock, auxiliary variable}

\title{\mytitle}

\author{Tatsuya Abe}

\date{STAIR Lab, Chiba Institute of Technology}

\begin{document}
\maketitle
\begin{abstract}
\myabstract
\end{abstract}
\textbf{Keywords:} \mykeywords

\allowdisplaybreaks

\newcommand{\myskip}{\vskip\baselineskip}

\def\seeappendix{
\begin{proof}
See Appendix.
\end{proof}
}
\newcommand{\gotoappendix}{
See Appendix in which we show the case analysis in detail, only for convenience of the reviews.}

\makeatletter
\newcommand\notsotiny{\@setfontsize\notsotiny\@vipt\@viipt}
\makeatother

\newcommand{\myverb}[1]{\texttt{\detokenize{#1}}}

\newcommand{\OK}{\mathord{\checkmark}}
\newcommand{\NA}{\mathord{\hspace{.6pt}\times\hspace{.6pt}}}

\newcommand{\letin}[2]{\ttlet \: #1 \: \texttt{in} \: #2}
\newcommand{\caseterm}[5]{\mathrm{case}(#1 , #2 . #3 , #4 . #5)}
\newcommand{\inl}[1]{\mathrm{inl}(#1)}
\newcommand{\inr}[1]{\mathrm{inr}(#1)}
\newcommand{\rec}[3]{\texttt{rec(}{#1\texttt{,}#2\texttt{,}#3}\texttt{)}}

\newcommand{\Value}{V}

\newcommand{\lamabs}[2]{\lambda #1 . #2}
\newcommand{\app}[2]{#1 #2}
\newcommand{\fst}[1]{\pi_0(#1)}
\newcommand{\snd}[1]{\pi_1(#1)}
\newcommand{\proj}[2]{\pi_{#1}(#2)}

\newcommand{\alphaequiv}{\equiv_\alpha}

\newcommand{\pair}[2]{\mathord{\langle #1,#2 \rangle}}
\newcommand{\triple}[3]{\mathord{\langle #1,#2,#3 \rangle}}
\newcommand{\quadruple}[4]{\mathord{\langle #1,#2,#3,#4 \rangle}}
\newcommand{\quintuple}[5]{\mathord{\langle #1,#2,#3,#4,#5 \rangle}}
\newcommand{\negop}[1]{\mathop{\neg}{#1}}
\newcommand{\neglit}[1]{\overline{#1}}
\newcommand{\existsop}[1]{\mathop{\exists}{#1}}
\newcommand{\forallop}[1]{\mathop{\forall}{#1}}
\newcommand{\boxop}[1]{\mathop{\Box}{#1}}
\newcommand{\diaop}[1]{\mathop{\textrm{\raisebox{-1pt}{\rotatebox{45}{\scalebox{0.8}{$\Box$}}}}}{#1}}
\newcommand{\boxopi}[2]{\mathop{\Box^#1}{#2}}
\newcommand{\diaopi}[2]{\mathop{\mathrm{\raisebox{-1pt}{\rotatebox{45}{\scalebox{0.8}{$\Box$}}}}^#1}{#2}}
\newcommand{\fv}[1]{\mathrm{fv}(#1)}
\newcommand{\ev}[2]{[\![#1]\!]_{#2}}
\newcommand{\evp}[2]{\langle\!|#1|\!\rangle_{#2}}
\newcommand{\dom}[1]{\mathrm{dom}(#1)}
\newcommand{\cod}[1]{\mathrm{cod}(#1)}
\newcommand{\tran}[1]{\mathrel{{#1}^{\mathord+}}}
\newcommand{\cls}[1]{\mathrel{{#1}^{\ast}}}
\newcommand{\lub}[1]{\lceil #1 \rceil}
\newcommand{\glb}[1]{\lfloor #1 \rfloor}
\newcommand{\intvl}[2]{[#1,#2)}
\newcommand{\intvlext}[4]{{\intvl{#3}{#4}}_{\langle #1,#2 \rangle}}
\newcommand{\oktyp}{\mathrm{ok}}
\newcommand{\size}[1]{\operatorname{size}(#1)}
\newcommand{\card}[1]{\#(#1)}
\newcommand{\norm}[1]{|\!|#1|\!|}

\newcommand{\add}[1]{\textcolor{red}{#1}}
\newcommand{\remove}[1]{\textcolor{cyan}{#1}}

\newcommand{\nolinefrac}[2]{\genfrac{}{}{0pt}{0}{#1}{#2}}

\newcommand{\manualproofenv}{\noindent\textit{\proofname}.\;\;}

\newcommand{\natvdots}{\smash{\vdots}\rule{0pt}{2ex}}
\newcommand{\manualqed}{\noindent\textit{\proofname}.\;\;}

\newcommand{\append}[2]{#1 \!\mathbin{\mbox{\raisebox{1pt}{\tiny $+$\!\!$+$}}}\! #2}

\newcommand{\judge}[3]{#1 \: \{\mathord{#2}\} \: #3}
\newcommand{\stable}[4]{#2 \: \{#3\} \Vdash^{#1} #4}
\newcommand{\unstable}[4]{#2 \not\Vdash^{#1} \{#3\} \; #4}

\newcommand{\assign}[3]{#2 \stackrel{\smash{#1}\rule{0pt}{3pt}}{\coloneqq} #3}
\newcommand{\nop}{\texttt{skip}}

\newcommand{\vt}[2]{#2\mathrm{@}#1}
\newcommand{\cons}[2]{#1 \mathbin{:} #2}
\newcommand{\tvl}{\mathit{L}}

\newcommand{\floors}[2]{\smash{\lfloor #1 \rfloor_{#2}}}
\newcommand{\jinv}[2]{\floors{#1}{#2}}
\newcommand{\suc}[1]{\underline{#1}}

\newcommand{\pre}{\mathit{B}}
\newcommand{\post}{\mathit{A}}
\newcommand{\rely}{\mathit{R}}
\newcommand{\guar}{\mathit{G}}
\newcommand{\delay}{\mathit{D}}

\newcommand{\comto}{\stackrel{\mathrm{c}}{\to}}
\newcommand{\envto}{\stackrel{\mathrm{e}}{\to}}
\newcommand{\delayto}{\stackrel{\mathrm{d}}{\to}}
\newcommand{\gento}[1]{\stackrel{#1}{\to}}

\newcommand{\ifend}[3]{\texttt{if}\:{#1}\mathord{?}{#2}\mathord{:}{#3}}
\newcommand{\whileend}[2]{\texttt{wl}\:{#1}\mathord{?}{#2}}

\lstset{language=C,
  frame=none,
  tabsize=4,
  basicstyle=\ttfamily,
  keywordstyle=\color{blue}\ttfamily,
  commentstyle=\ttfamily,%\color{purple}\ttfamily,
  columns=[l]{fullflexible},
%  numbers=left,
  numberstyle=\scriptsize,
  stepnumber=1,
  numbersep=5pt,
  escapechar=\#,
  keepspaces=true
}

\section{Introduction}\label{sec:intro}

Modern computer architectures have multiple cores. Because using
multiple cores is significant in computational performance, we cannot
ignore weak memory models. On a weak memory model, delays in the
effects of store instructions are allowed. For example, let us
consider the example program called Store Buffering:

\lstset{frame=none}
{\centering
\small
\begin{minipage}{75pt}
\begin{lstlisting}
 x := 1
 r0 := y  // 0
\end{lstlisting}
\end{minipage}
\quad
\lstset{frame=L}
\begin{minipage}{70pt}
\begin{lstlisting}
 y := 1
 r1 := x  // 0
\end{lstlisting}
\end{minipage}\par
}

\noindent
where $x$ and $y$ are shared variables and $r_0$ and $r_1$ are
thread-local variables.

On the strongest memory model called strict consistency, $r_0 = 0
\wedge r_1 = 0$ does not finally hold. If $r_0 = 0$ holds,
$\assign{}{r_0}{y}$ is executed earlier than $\assign{}{y}{1}$. It
implies that $\assign{}{x}{1}$ is executed earlier than
$\assign{}{r_1}{x}$ and therefore $r_1 = 1$ holds, and vice versa. On
a weak memory model that allows delays in the effects of store
instructions, $r_0 = 0 \wedge r_1 = 0$ is allowed.

Because the example program has just six execution traces on strict
consistency, the so-called exhaustive search enables us to confirm
that $r_0 = 0 \wedge r_1 = 0$ does not finally hold.  Even if we
cannot assume strict consistency, model checking with weak memory
models is still a promising verification method
(e.g.,~\cite{abdulla2016,koko2021}). However, we would like to focus
on program logic sound to weak memory models in this paper.

Owicki--Gries logic is well-known to be sound to the standard state
semantics based on strict consistency~\cite{Owicki1976A}. The logic is
an extension for concurrent programs of Hoare logic for
sequential programs~\cite{DBLP:journals/cacm/Hoare69}. Owicki and
Gries invented the notion of non-interference to repair the
unsoundness of the parallel composition rule without any side
condition. They also invented the notion of auxiliary variables to
enhance the provability of their logic.

Because Owicki--Gries logic fits strict consistency, it is too strong
to be sound to weak memory models. In this paper, we modify the
non-interference and propose program logic sound to a weak memory model that
allows delays in the effects of store instructions.

In this paper, we adopt timestamp semantics as a weak memory
model. Timestamp semantics are obtained by extending messages on
memory to messages with timestamps and letting threads have the
so-called vector clock, which has one clock at each shared
variable~\cite{Lamport78,mattern1989}. It is known that it expresses
delays in the effects of store instructions.  For example, because the
store instructions that occur in Store Buffering send the messages
$x=1$ and $y=1$ with timestamp $1$ to the memory as follows:

\lstset{frame=none}
{\centering
\small
\begin{minipage}{110pt}
\begin{lstlisting}
 x := 1   // <<x,1>,1>
 r0 := y  // <<y,0>,0>
\end{lstlisting}
\end{minipage}
\quad
\lstset{frame=L}
\begin{minipage}{110pt}
\begin{lstlisting}
 y := 1   // <<y,1>,1>
 r1 := x  // <<x,0>,0>
\end{lstlisting}
\end{minipage}\par
}

\noindent
the load instructions read the messages $x=0$ and $y=0$ with timestamp
$0$. Therefore, $r_0 = 0 \wedge r_1 =0$ is allowed to hold finally.
On the other hand, in a similar example program called Coherence:

\lstset{frame=none}
{\centering
\small
\begin{minipage}{110pt}
\begin{lstlisting}
 x := 1   // <<x,1>,1>
 r0 := x  // <<x,2>,2>
\end{lstlisting}
\end{minipage}
%\;\makebox[0pt][c]{\raisebox{15pt}{$\xymatrix@C=160pt{\ar@{-}[rd]&\ar@{-}[ld]\\&}$}}
\quad
\lstset{frame=L}
\begin{minipage}{110pt}
\begin{lstlisting}
 x := 2   // <<x,2>,2>
 r1 := x  // <<x,2>,2>
\end{lstlisting}
\end{minipage}

}

\noindent
$r_0 = 2 \wedge r_1 = 1$ is not allowed to finally hold because the
memory requires a total order of messages with timestamps for the same
shared variable.

In this paper, we modify the non-interference in Owicki--Gries logic
and construct concurrent program logic sound to timestamp
semantics. The key idea is to formalize different observations of
shared variables by threads. Also, this study provides three
theoretically interesting techniques clarified by the novel
non-interference.  The author believes that broadly sharing the
techniques contributes to a better understanding of constructing
logic. The techniques are explained in detail in
Section~\ref{sec:ours} and clearly described in
Section~\ref{sec:conclusion} that concludes the paper.

\section{Owicki--Gries Logic}\label{sec:og}

We introduce notation for describing the standard state semantics for
concurrent programs and remember Owicki--Gries logic, which is sound
to the semantics.

Expression $E$ and sequential program $s$ are defined as follows:
\begin{align*}
  E & \Coloneqq n \mid v \mid E + E \mid E - E \mid \ldots &
  s &  \Coloneqq \nop \mid \assign{}{v}{E} \mid s; s
\end{align*}
where $n$ and $v$ denote a numeral and a variable,
respectively. Expressions denoting arithmetic operations are
appropriately defined. No operation $\nop$ is used only to define
operational semantics. We assume that $\nop$ does not occur in user
programs.

In this paper, we do not adopt the so-called fork-join paradigm for
parallel executions but fix the number of threads to $N$. A concurrent
program is denoted by
\[
s_0 \parallel \cdots \parallel s_i \parallel \cdots \parallel s_{N-1} \enspace .
\]
We also omit conditional and loop statements for the simplicity of
the presentation.

State $M$ is a function from variables to numerals in a standard
manner. Configurations consist of pairs $\pair{s}{M}$ of programs and
states. We define operational semantics as follows:
\begin{center}
    \begin{tabular}{c@{\qquad}c}
      $\dfrac{
      }
      {\pair{\assign{}{v}{E}}{M} \to \pair{\nop}{M [v \mapsto \ev{E}{M}]}}
      $
      &
      $\dfrac{
        \pair{s_0}{M} \to \pair{\nop}{M'}
      }
      {\pair{s_0; s_1}{M} \to \pair{s_1}{M'}}
      $
      \\
      \\
      $\dfrac{
        \pair{s_0}{M} \to \pair{s'_0}{M'}
      }
      {\pair{s_0; s_1}{M} \to \pair{s'_0; s_1}{M'}}
      $
      &
      $\dfrac{
        \pair{s_i}{M} \to \pair{s'_i}{M'}
      }
      {\nolinefrac{\pair{s_0 \parallel \cdots \parallel s_i \parallel \cdots \parallel s_{N-1}}{M} \quad\;\;}{\to \pair{s_0 \parallel \cdots \parallel s'_i \parallel \cdots \parallel s_{N-1}}{M'}}}
      $
    \end{tabular}
\end{center}
where $\ev{E}{M}$ means the interpretation $E$ by $M$ that is defined as follows:
\begin{align*}
  \ev{n}{M} & = n &
  \ev{v}{M} & = M(v) &
  \ev{E + E'}{M} & = \ev{E}{M} + \ev{E'}{M} \enspace \ldots
\end{align*}
and the update function $M [v \mapsto n] (v')$ is defined as follows:
\[
M [v \mapsto n] (v') = \begin{cases}n & \mbox{if } v' = v\\M(v') & \mbox{otherwise.}\end{cases}
\]

To describe properties that concurrent programs enjoy, we define an
assertion language as follows:
\begin{align*}
  \varphi & \Coloneqq  E = E \mid E \leq E \mid% \top \mid \bot \mid
  \negop{\varphi} \mid \varphi \wedge \varphi \enspace .
\end{align*}
We use abbreviations $\top$, $\varphi \vee \psi$, and $\varphi \supset
\psi$ in the standard manner.

We interpret $\varphi$ by $M$ as follows:
\begin{align*}
  M \vDash E = E'
  & \Longleftrightarrow
  \ev{E}{M} = \ev{E'}{M} &
  M \vDash E \leq E'
  & \Longleftrightarrow
  \ev{E}{M} \leq \ev{E'}{M}\\
  M \vDash \negop{\varphi}
  & \Longleftrightarrow
  M \not\vDash \varphi &
  M \vDash \varphi \wedge \psi
  & \Longleftrightarrow
  M \vDash \varphi \mbox{ and }
  M \vDash \psi
  \enspace .
\end{align*}

Owicki and Gries provided logic that is sound to the state
semantics~\cite{Owicki1976A}. Owicki--Gries logic consists of the
so-called Hoare triples. A triple $\judge{\varphi}{s}{\psi}$ means
that an execution of the program $s$ under the pre-condition $\varphi$
implies that the post-condition $\psi$ holds.

The following is a subset of their inference rules:
\begin{center}
    \begin{tabular}{cc}
      %% $\dfrac{
      %% }
      %% {\judge{\varphi}{\nop}{\varphi}}
      %% $
      $\dfrac{
      }
      {\judge{[E/v]\varphi}{\assign{}{v}{E}}{\varphi}}
      $
      &
      $\dfrac{
          \vDash \varphi' \supset \varphi\qquad
          \judge{\varphi}{s}{\psi}\qquad
          \vDash \psi \supset \psi'
      }
      {\judge{\varphi'}{s}{\psi'}}
      $
      \\
      \\
      $\dfrac{
          \judge{\varphi}{s}{\psi}\qquad
          \judge{\psi}{s'}{\chi}
      }
      {\judge{\varphi}{s; s'}{\chi}}
      $
      &
      {
      $\dfrac{
          \nolinefrac{\varDelta_i}{\judge{\varphi_i}{s_i}{\psi_i}}\qquad
          \nolinefrac{}{\mbox{$\varDelta_i$s are non-interfering}}
      }
      {\judge{\varphi_0 \wedge \cdots \wedge \varphi_{N-1}}{s_0 \parallel \cdots \parallel s_{N-1}}{\psi_0 \wedge \cdots \wedge \psi_{N-1}}}
      \enspace .
      $
      }
    \end{tabular}
\end{center}

Formula $[E/v]\varphi$ in the assignment axiom denotes the formula
obtained by replacing $v$ that occurs in $\varphi$ by $E$.
Symbol $\vDash$ occurring in the third inference rule denotes the
standard satisfaction relation as seen in the first-order predicate
logic.
If there exists a derivation tree of the root $\judge{\varphi}{s_0
  \parallel \cdots \parallel s_{N-1}}{\psi}$, we write $\vdash
\judge{\varphi}{s_0 \parallel \cdots \parallel s_{N-1}}{\psi}$.

Owicki and Gries defined that $\varDelta_1$ does not interfere with
$\varDelta_0$ if $\vdash \judge{\varphi_1 \wedge
  \varphi_0}{\assign{}{v_1}{E_1}}{\varphi_0}$ and $\vdash
\judge{\varphi_1 \wedge \psi_0}{\assign{}{v_1}{E_1}}{\psi_0}$ hold for
any $\judge{\varphi_0}{\assign{}{v_0}{E_0}}{\psi_0}$ in $\varDelta_0$
and $\judge{\varphi_1}{\assign{}{v_1}{E_1}}{\_}$ in $\varDelta_1$. We
often write $\_$ for a placeholder. In
this paper, we write
$\stable{}{\varphi_1}{\assign{}{v_0}{E_1}}{\varphi_0}$ for $\vdash
\judge{\varphi_1 \wedge \varphi_0}{\assign{}{v_1}{E_1}}{\varphi_0}$,
for short, and call that $\varphi_0$ is stable under
$\stable{}{\varphi_1}{\assign{}{v_1}{E_1}}{\_}$.  We call
$\varDelta_i$s $(0 \leq i < N)$ non-interfering if $\varDelta_{i_0}$
does not interfere with $\varDelta_{i_1}$ for any $i_0 \neq i_1$.

\begin{figure}
\input{confluence.tex}
\caption{Parallel compositionality by non-interference.}\label{fig:confluence}
\end{figure}

In a standard manner, we define $\vDash \judge{\varphi}{s}{\psi}$ if
$M \vDash \varphi$ and $\pair{s}{M} \cls{\to} \pair{\nop}{M'}$ implies
$M'\vDash \psi$ for any $M$ and $M'$ where $\cls{\to}$ is the
reflexive and transitive closure of $\to$.
Similarly, we define $\vDash \judge{\varphi}{s_0 \parallel \cdots
  \parallel s_{N-1}}{\psi}$.

Figure~\ref{fig:confluence} shows an example that the non-interference
implies a kind of confluence denoting that the conjunction $\varphi_0
\wedge \varphi_1$ of the first pre-conditions implies the conjunction
$\chi_0 \wedge \chi_1$ of the last post-conditions, where $\varphi
\xrightarrow{\smash{\{s\}}} \psi$ denotes $\judge{\varphi}{s}{\psi}$.
Thus, the logic is sound to the conventional state semantics.

Let us see an example derivation. In this paper, we assume that
initial values of variables are $0$.
Figure~\ref{fig:ch_OG} shows a derivation of Coherence for ensuring
the property $\negop{(r_0 = 2 \wedge r_1 = 1)}$.
In this paper, we write $x$ and $y$ for shared variables and write $r$
for a thread-local variable.

\begin{figure}
\input{ch_OG.tex}
\caption{Derivation of Coherence on non-interference.}\label{fig:ch_OG}
\end{figure}

Notably, the shared variable $x$ acts as an intermediary
between $r_0$ and $r_1$ and ensures finally $r_0 \leq r_1$ as shown in
the derivation. Such variables often enable us to describe
assertions.

%We implemented all the derivations described in this paper on Horn-ICE Verification Toolkit~\cite{DBLP:journals/pacmpl/EzudheenND0M18}.

We checked all the derivations described in this paper using the Boogie
program verifier~\cite{barnett2005boogie}.

The provability of the logic still needs to be
improved. Figure~\ref{fig:onetwo_OG} shows the concurrent program
$\assign{}{x}{x+1} \parallel \assign{}{x}{x+2}$ and a derivation for
ensuring that $x=3$ finally holds. A reason that we can construct the
derivation is that the assertions have sufficient information about
the execution traces of the program. Specifically, the value of $x$
tells us where the program is executed.

\begin{figure}
\newcommand{\oneone}{x = 0 \vee x = 2}
\newcommand{\onecomone}{\assign{}{x}{x + 1}}
\newcommand{\onetwo}{x = 1 \vee x = 3}
\newcommand{\twoone}{x = 0 \vee x = 1}
\newcommand{\twocomone}{\assign{}{x}{x + 2}}
\newcommand{\twotwo}{x = 2 \vee x = 3}
{\small
\lstset{frame=none}
\begin{minipage}{.20\textwidth}
\begin{lstlisting}
 x := x + 1
\end{lstlisting}
\end{minipage}
%\;\makebox[0pt][c]{\raisebox{15pt}{$\xymatrix@C=160pt{\ar@{-}[rd]&\ar@{-}[ld]\\&}$}}
\lstset{frame=L}
\begin{minipage}{.25\textwidth}
\begin{lstlisting}
 x := x + 2
\end{lstlisting}
\end{minipage}
}
{\footnotesize
\begin{minipage}{.20\textwidth}
  \[
  \begin{array}{c}
    \smash{\oneone}\rule{0pt}{2.5ex}\\
    \smash{\{\onecomone\}}\rule{0pt}{2.5ex}\\
    \smash{\onetwo}\rule{0pt}{2.5ex}
  \end{array}
  \]
\end{minipage}
\begin{minipage}{.20\textwidth}
  \[
  \begin{array}{c}
    \smash{\twoone}\rule{0pt}{2.5ex}\\
    \smash{\{\twocomone\}}\rule{0pt}{2.5ex}\\
    \smash{\twotwo}\rule{0pt}{2.5ex}
  \end{array}
  \]
\end{minipage}
\vspace{\baselineskip}

invariant: $x = 0 \vee x = 1 \vee x = 2 \vee x = 3$\\
}

  {\footnotesize
\begin{minipage}{.52\textwidth}
$\stable{}{\twoone}{\twocomone}{\oneone}$

$\stable{}{\twoone}{\twocomone}{\onetwo}$
\end{minipage}
\begin{minipage}{.55\textwidth}
$\stable{}{\oneone}{\onecomone}{\twoone}$

$\stable{}{\oneone}{\onecomone}{\twotwo}$
\end{minipage}
}
\caption{Derivation of One-Two on non-interference.}\label{fig:onetwo_OG}
\end{figure}

On the other hand, the value of $x$ in the parallel composition
$\assign{}{x}{x+1} \parallel \assign{}{x}{x+1}$ does not tell
us. Therefore, we cannot add appropriate assertions that do not
interfere with each other to the as-is program.

\begin{figure}
\input{oneone_OG.tex}
\caption{Derivation of One-One on non-interference.}\label{fig:oneone_OG}
\end{figure}

Owicki and Gries invented auxiliary variables, which are write-only in
programs, atomically updated with existing assignments, described in
assertions, and removed from programs
later. Figure~\ref{fig:oneone_OG} shows a derivation for ensuring that
$x=2$ finally holds. The auxiliary variables $a_0$ and $a_1$ enable us
to describe the assertions.

\begin{theorem}[\cite{Owicki1976A}]\label{thm:og}
  $\vdash \judge{\varphi}{s_0 \parallel \cdots \parallel s_{N-1}}{\psi}$ implies
  $\vDash \judge{\varphi}{s_0 \parallel \cdots \parallel s_{N-1}}{\psi}$.
\end{theorem}

Figure~\ref{fig:sb_OG} shows a derivation of Store Buffering with
auxiliary variables.

\begin{figure}
\newcommand{\oneone}{\top}
\newcommand{\onetwo}{x = 1}
\newcommand{\onethree}{a_0 = 1}
\newcommand{\onecomone}{\assign{}{x}{1}}
\newcommand{\onecomtwo}{\assign{}{r_0}{y}, \assign{}{a_0}{1}}
\newcommand{\twoone}{a_0 \leq x}
\newcommand{\twotwo}{y = 1 \wedge a_0 \leq x}
\newcommand{\twothree}{y = 1 \wedge (a_0 \leq r_0 \vee r_1 = 1)}
\newcommand{\twocomone}{\assign{}{y}{1}}
\newcommand{\twocomtwo}{\assign{}{r_1}{x}}
{\small
\lstset{frame=none}
\begin{minipage}{.17\textwidth}
\begin{lstlisting}
 x := 1
 atomic {
  r0 := y
  a0 := 1
 }
\end{lstlisting}
\end{minipage}
\lstset{frame=L}
\begin{minipage}{.17\textwidth}
\begin{lstlisting}
 y := 1
 atomic {
  r1 := x
 }
 #\phantom{X}#
\end{lstlisting}
\end{minipage}
}
{\footnotesize
\begin{minipage}{.19\textwidth}
  \[
  \begin{array}{c}
    \smash{\oneone}\rule{0pt}{2.5ex}\\
    \smash{\{\onecomone\}}\rule{0pt}{2.5ex}\\
    \smash{\onetwo}\rule{0pt}{2.5ex}\\
    \smash{\{\onecomtwo\}}\rule{0pt}{2.5ex}\\
    \smash{\onethree}\rule{0pt}{2.5ex}
  \end{array}
  \]
\end{minipage}
\qquad
\begin{minipage}{.29\textwidth}
  \[
  \begin{array}{c}
    \smash{\twoone}\rule{0pt}{2.5ex}\\
    \smash{\{\twocomone\}}\rule{0pt}{2.5ex}\\
    \smash{\twotwo}\rule{0pt}{2.5ex}\\
    \smash{\{\twocomtwo\}}\rule{0pt}{2.5ex}\\
    \smash{\twothree}\rule{0pt}{2.5ex}
  \end{array}
  \]
\end{minipage}
\vspace{\baselineskip}

invariant: $(x = 0 \vee x = 1) \wedge (y = 0 \vee y = 1) \wedge (r_0 = 0 \vee r_0 = 1) \wedge (r_1 = 0 \vee r_1 = 1) \wedge (a_0 = 0 \vee a_0 = 1)$\\

}
{\footnotesize
\begin{minipage}{.45\textwidth}
$\stable{}{\twoone}{\twocomone}{\oneone}$

$\stable{}{\twoone}{\twocomone}{\onetwo}$

$\stable{}{\twoone}{\twocomone}{\onethree}$

$\stable{}{\twotwo}{\twocomtwo}{\oneone}$

$\stable{}{\twotwo}{\twocomtwo}{\onetwo}$

$\stable{}{\twotwo}{\twocomtwo}{\onethree}$
\end{minipage}
\begin{minipage}{.54\textwidth}
$\stable{}{\oneone}{\onecomone}{\twoone}$

$\stable{}{\oneone}{\onecomone}{\twotwo}$

$\stable{}{\oneone}{\onecomone}{\twothree}$

$\stable{}{\onetwo}{\onecomtwo}{\twoone}$

$\stable{}{\onetwo}{\onecomtwo}{\twotwo}$

$\stable{}{\onetwo}{\onecomtwo}{\twothree}$
\end{minipage}
}
\caption{Derivation of Store Buffering on non-interference.}\label{fig:sb_OG}
\end{figure}

\section{Timestamp Semantics}

We formally define timestamp semantics.

In the following, we formally distinguish thread-local variables $r$
from shared variables $x$. We also divide assignments into three
kinds, thread-local, load, and store instructions as follows:
\begin{align*}
  e & \Coloneqq n \mid r \mid e + e \mid \ldots &
  c &  \Coloneqq \assign{i}{r}{e} \mid \assign{i}{r}{x} \mid \assign{i}{x}{e} &
  s &  \Coloneqq \nop \mid c \mid s; s
\end{align*}
and each assignment has a thread identifier denoting the thread that
executes the assignment.  A sequentially compound statement contains a
unique thread identifier.

On timestamp semantics, thread-local states $S$ are separated from the
shared memory $M$. Thread-local state $S$ is a function from
thread-local variables to numerals. We define the interpretations of
expressions by $S$ as follows:
\begin{align*}
  \ev{n}{S} & = n & \ev{r}{S} & = S(r) & \ev{e + e'}{S} & =
  \ev{e}{S} + \ev{e'}{S} \enspace \ldots \enspace .
  %\mid \ifend{\varphi}{c}{c} \mid \whileend{\varphi}{c}
\end{align*}

The shared memory possesses messages with timestamps
$\pair{\pair{x}{t}}{n}$, differently from physical memories that we
know. It is known that a set of timestamps should be dense
(e.g.~\cite{conf/popl/Kang2017}).  The shared memory $M$ is a function
from pairs of shared variables and timestamps to numerals.  Notably,
$M$ is a function; a shared memory cannot possess multiple messages
for the same shared variable and the same timestamp.

Each thread has a vector clock~\cite{Lamport78,mattern1989}.  A vector
clock takes a shared variable and returns a timestamp. Differences
between vector clocks of threads express observations of shared
variables by threads.

Thread-local configurations $C$ consist of triples $\triple{s}{S}{T}$
of programs, thread-local states, and vector clocks.  Thread-local
configurations with memory consist of pairs $\pair{C}{M}$ of
thread-local configurations $C$ and memories.  Global Configurations
consist of pairs $\pair{\mathfrak{C}}{M}$ of thread-local
configurations $\mathfrak{C}$ and memories $M$ where $\mathfrak{C}$
takes a thread identifier $i$ and returns a configuration $C$.
Operational semantics are as follows:
\begin{center}
    \begin{tabular}{c}
      $\dfrac{
        \pair{\mathfrak{C}(i)}{M} \to \pair{C'}{M'}
      }
      {\pair{\mathfrak{C}}{M} \to \pair{\mathfrak{C}[i \mapsto C']}{M'}}
      $
\quad
      $\dfrac{
      }
      {\pair{\triple{\assign{i}{r}{e}}{S}{T}}{M} \to\pair{\triple{\nop}{S [r \mapsto \ev{e}{S}]}{T}}{M}}
      $
      \\
      \\
      $\dfrac{
        M(\pair{x}{t}) = n \qquad T(x) \leq t
      }
    {\pair{\triple{\assign{i}{r}{x}}{S}{T}}{M} \to \pair{\triple{\nop}{S [r \mapsto n]}{T[x \mapsto t]}}{M}}
      $
      \\
      \\
      $\dfrac{
        T(x) < t
      }
      {\pair{\triple{\assign{i}{x}{e}}{S}{T}}{M} \to \pair{\triple{\nop}{S}{T[x \mapsto t]}}{M \sqcup \{\pair{\pair{x}{t}}{\ev{e}{S}}\}}}
      $
      \\
      \\
      $\dfrac{
        \pair{\triple{s_0}{S}{T}}{M} \to\pair{\triple{\nop}{S'}{T'}}{M'}
      }
      {\pair{\triple{s_0; s_1}{S}{T}}{M} \to\pair{\triple{s_1}{S'}{T'}}{M'}}
      $
      \;
      $\dfrac{
        \pair{\triple{s_0}{S}{T}}{M} \to\pair{\triple{s'_0}{S'}{T'}}{M'}
      }
      {\pair{\triple{s_0; s_1}{S}{T}}{M} \to\pair{\triple{s'_0; s_1}{S'}{T'}}{M'}}
       \;.$
    \end{tabular}
\end{center}

No explanation is needed for thread-local assignments. An assignment
of a load instruction takes any message in the memory whose timestamp
is equal to or larger than the timestamp that the vector clock
indicates.  Furthermore, it updates the vector clock.
An assignment of a store instruction sends a message whose timestamp
is strictly larger than the timestamp that the vector clock indicates,
and updates the vector clock.
Timestamp semantics express a modern memory model on which effects
of store instructions may be delayed (e.g.~\cite{conf/popl/Kang2017}).

\section{Observation-Based Logic}\label{sec:ours}

In this section, we provide concurrent program logic for timestamp
semantics.

\subsection{Formal system}

We introduce observation
variable $x^i$, which denotes $x$ observed by thread $i$, for any
shared variable $x$. Whereas the notion of observation variables is
previously provided by the author~\cite{am2016aplas}, we provide
another logic in this paper.

The motivation for the introduction of observation variables is
straightforward.  In the standard state semantics, all variables are
synchronous; if a thread updates the value of a variable, then other
threads can immediately see the update. Variables that occur in
assertions are also synchronous in stability checking.
However, in timestamp semantics, that is not the case. Nevertheless,
variables that occur in assertions in logic should be synchronous
because stability checking is designed that way.

Observation variables fix the gap. Observation variables are
synchronous in assertions in logic, that is, if a thread updates the
value of an observation variable, then other threads can immediately
see the update.
Surely, a relation between observation variables $x^{i_0}$ and
$x^{i_1} \; (i_0 \neq i_1)$ is not a relation between distinct shared
variables $x$ and $y$. We clarify and formalize the relation between
observation variables in logic.

We also introduce timestamp variable $\vt{x}{t}$ for any shared
variable $x$. We describe how to use timestamp variables with the
explanation of non-observation-interference described later. Timestamp
variables in assertions are interpreted as thread-local variables.
The following are inference rules consisting of extended judgments:
\begin{center}
    \begin{tabular}{c@{\qquad}c}
      %% $\dfrac{}{\judge{\varphi, \tvl}{\nop}{\varphi, \tvl}}
      %% $
      $\dfrac{}{\judge{[e/r]\varphi, \tvl}{\assign{i}{r}{e}}{\varphi, \tvl}}
      $
      &
      $\dfrac{}{\judge{[x^i/r]\varphi, \tvl}{\assign{i}{r}{x}}{\varphi, \tvl}}
      $
      \\
      \\
      $\dfrac{
        \mbox{
          $t$ is fresh
        }
      }
      {\judge{[e/x^i]\varphi, \tvl}{\assign{i}{x}{e}}{\varphi, \tvl[x \mapsto \cons{\tvl(x)}{\vt{x}{t}}]}}
      $
      &
      $\dfrac{
          \judge{\varphi, \tvl_0}{s_0}{\psi, \tvl}\qquad
          \judge{\psi, \tvl}{s_1}{\chi, \tvl_1}
      }
      {\judge{\varphi, \tvl_0}{s_0; s_1}{\chi, \tvl_1}}
      $
      \\
      \\
      \multicolumn{2}{c}{
      $\dfrac{
          \vDash \varphi' \supset \varphi\qquad
          \judge{\varphi, \tvl_0}{s}{\psi, \tvl_1}\qquad
          \vDash \psi \supset \psi'
      }
      {\judge{\varphi', \tvl_0}{s}{\psi', \tvl_1}}
      $
}
      \\
      \multicolumn{2}{c}{
      $\dfrac{
          \nolinefrac{\varDelta_i}{\judge{\varphi_i, \tvl_\varnothing}{s_i}{\psi_i, \tvl_i}}\qquad
          \nolinefrac{}{\mbox{$\varDelta_i$s are non-observation-interfering}}
      }
      {\judge{\varphi_0 \wedge \cdots \wedge \varphi_{N-1}}{s_0 \parallel \cdots \parallel s_{N-1}}{\psi_0 \wedge \cdots \wedge \psi_{N-1}}, \tvl_0, \ldots, \tvl_{N-1}} \enspace .
      $
      }
    \end{tabular}
\end{center}

Because assignments are divided into three kinds, the assignment axiom
is also divided into three. Assignments of load instructions by thread
$i$ replace thread-local variables with $x^i$. Careful readers may
wonder if the assignment axiom for load instructions is
unconditionally unsound. The unsoundness is remedied by the
non-observation-interference, as explained in detail in
Section~\ref{sec:nbi}.

The assignment axiom for store instructions is sound because we
introduced observation variables to make it so.  For example,
$\judge{r + 1 = 2 \wedge x^{i_1} = 0}{\assign{i_0}{x}{r + 1}}{x^{i_0}
  = 2 \wedge x^{i_1} = 0}$ is valid because $\assign{i_0}{x}{r + 1}$
updates the vector clock of thread $i_0$, and does not update the
vector clock of thread $i_1$, that is, $x^{i_1} = 0$ is kept without
any loading by thread $i_1$.

Sequence $\tvl$ consists of reverse lists of timestamp variables at each
shared variable. An initial timestamp sequence $\tvl_\varnothing$ has exactly one
timestamp variable for any shared variable. A timestamp variable is
freshly generated at an assignment axiom of a store instruction to $x$
and is added to the end at $x$ of the timestamp sequence.
To ensure a total order of timestamps of each shared variable,
we assume $\vdash^t \vt{x}{t} < \vt{x}{t'}$ for any shared variable $x$
and
$\cons{\ldots}{\cons{\vt{x}{t}}{\cons{\ldots}{\cons{\vt{x}{t'}}{\ldots}}}}$
in $\tvl$.
To ensure a total order of timestamps of each shared variable on
memory, we assume $\vdash^t \vt{x}{t} < \vt{x}{t'} \vee \vt{x}{t'} <
\vt{x}{t}$ for any shared variable $x$ and any distinct $t$ and $t'$.
If there exists a derivation tree of the root
$\judge{\varphi, \tvl}{s}{\psi, \tvl'}$ in the system, we write $\vdash^t
\judge{\varphi, \tvl}{s}{\psi, \tvl'}$. Similarly, we define $\vdash^t
\judge{\varphi}{s_0 \parallel \cdots \parallel s_{N-1}}{\psi, \tvl_0, \ldots, \tvl_{N-1}}$.

We define a satisfaction relation that configurations enjoy
assertions.
For convenience, we write $\mathfrak{S}$ for the thread-local states
of $\mathfrak{C}$, that is, $\mathfrak{S}(i)$ denotes the thread-local
state of $\mathfrak{C}(i)$. We also write $\mathfrak{T}$ for the
vector clocks of $\mathfrak{C}$, that is, $\mathfrak{T}(i)$ denotes
the vector clock of $\mathfrak{C}(i)$.
We write $\rho$ for a function from timestamp variables to
values.
We define $\pair{M}{\pair{\mathfrak{S}}{\mathfrak{T}}}, \rho \vDash^t \varphi$
as follows:
\begin{align*}
  E  \Coloneqq e & \mid x^i \mid \vt{x}{t} \mid E + E \mid E - E \mid \ldots
\end{align*}
\vspace{-2\baselineskip}
\begin{align*}
  \ev{n}{\pair{M}{\pair{\mathfrak{S}}{\mathfrak{T}}}, \rho} & = n\\
  \ev{r}{\pair{M}{\pair{\mathfrak{S}}{\mathfrak{T}}}, \rho} & = \mathfrak{S}(i)(r) \mbox{ if thread $i$ contains $r$}\\
  \ev{x^i}{\pair{M}{\pair{\mathfrak{S}}{\mathfrak{T}}}, \rho} & = M(\pair{x}{\mathfrak{T}(i)(x)})\\
  \ev{\vt{x}{t}}{\pair{M}{\pair{\mathfrak{S}}{\mathfrak{T}}}, \rho} & = \rho(\vt{x}{t})\\
  \ev{E + E'}{\pair{M}{\pair{\mathfrak{S}}{\mathfrak{T}}}, \rho} & = \ev{E}{\pair{M}{\pair{\mathfrak{S}}{\mathfrak{T}}}, \rho} + \ev{E'}{\pair{M}{\pair{\mathfrak{S}}{\mathfrak{T}}}, \rho} \enspace \ldots
\end{align*}
\vspace{-2\baselineskip}
\begin{align*}
  \pair{M}{\pair{\mathfrak{S}}{\mathfrak{T}}}, \rho \vDash^t E = E'
  & \Longleftrightarrow
  \ev{E}{\pair{M}{\pair{\mathfrak{S}}{\mathfrak{T}}}, \rho} = \ev{E'}{\pair{M}{\pair{\mathfrak{S}}{\mathfrak{T}}}, \rho}\\
  \pair{M}{\pair{\mathfrak{S}}{\mathfrak{T}}}, \rho \vDash^t E \leq E'
  & \Longleftrightarrow
  \ev{E}{\pair{M}{\pair{\mathfrak{S}}{\mathfrak{T}}}, \rho} = \ev{E'}{\pair{M}{\pair{\mathfrak{S}}{\mathfrak{T}}}, \rho}\\
  \pair{M}{\pair{\mathfrak{S}}{\mathfrak{T}}}, \rho \vDash^t \negop{\varphi}
  & \Longleftrightarrow
  \pair{M}{\pair{\mathfrak{S}}{\mathfrak{T}}}, \rho \not{\vDash^t} \varphi\\
  \pair{M}{\pair{\mathfrak{S}}{\mathfrak{T}}}, \rho \vDash^t \varphi \wedge \psi
  & \Longleftrightarrow
  \pair{M}{\pair{\mathfrak{S}}{\mathfrak{T}}}, \rho \vDash^t \varphi \mbox{ and }
  \pair{M}{\pair{\mathfrak{S}}{\mathfrak{T}}}, \rho \vDash^t \psi \enspace .
\end{align*}

We say that $\rho$ follows $\tvl$ if for any shared variable $x$,
the timestamp variables about $x$ in $\tvl$ are increasing with
respect to $\rho$-values.
For example, $\rho$ follows $\{ [\vt{x}{t_1},
  \vt{x}{t_2}, \vt{x}{t_3}],$ $[\vt{y}{t_1}, \vt{y}{t_2}] \}$ where
$\rho (\vt{x}{t_j}) = j$ and $\rho (\vt{y}{t_j}) = j$.

We call a sequence of transitions $\pair{\mathfrak{C}_0}{M_0}
\cls{\to} \pair{\mathfrak{C}_m}{M_n}$ consistent with $\rho$ if for
any $\vt{x}{t}$ freshly genenerated by a store instruction, the
message in the memories passed by the store instruction is
$\pair{\pair{x}{\rho(\vt{x}{t})}}{\_}$.

We assume $\rho$ follows $\tvl$. We define
$\pair{M}{\pair{\mathfrak{S}}{\mathfrak{T}}},\rho \vDash^t
\judge{\varphi, \tvl_0}{s_0 \parallel \cdots \parallel s_{N-1}}{\psi,
  \tvl_1}$ if $\pair{M}{\pair{\mathfrak{S}}{\mathfrak{T}}}, \rho
\vDash^t \varphi$ and
$\pair{\triple{s_0 \parallel \cdots \parallel s_{N-1}}{\mathfrak{S}(i)}{\mathfrak{T}(i)}}{M} \cls{\to}
\pair{\triple{\nop}{\mathfrak{S'}(i)}{\mathfrak{T'}(i)}}{M'}$
consistent with $\rho$ imply
$\pair{M'}{\pair{\mathfrak{S'}}{\mathfrak{T'}}}, \rho \vDash^t \psi$
for any $\mathfrak{S}$ and $\mathfrak{T}$.

\subsection{Non-observation-interference}\label{sec:nbi}

Before providing the formal definition of
non-observation-interference, we explain its intuition.
For simplicity, we often omit reverse lists of timestamp variables
in judgments if they are clear from the context.

First, it appears that $\judge{x^{i_0} = 1}{\assign{i_0}{r}{x}}{r = 1
  \wedge x^{i_0} = 1}$ is unsound to timestamp semantics because the
value of $x$ at a timestamp $t$, which may not be pointed by the
vector clock satisfying $x^{i_0}=1$, is loaded. That makes us want to
add, for example, a necessity modality operator $\Box$ meaning
``always true'' to $x^{i_0}=1$ in the pre-condition. Alternatively,
one may want to extend the assertion language and modify the
interpretation of Hoare triples~\cite{DDDW20}. However, we do not
adopt them in this paper.

If the number of threads is $1$, the judgment is still sound because
the memory has no message from another thread. Soundness of
$\judge{\varphi_0}{\assign{i_0}{r}{x}}{\psi_0}$ is stained by a store
instruction $\assign{i_1}{x}{e}$ by another thread $i_1$. It is
reminiscent of the non-interference.

Let $\judge{\varphi_1}{\assign{i_1}{x}{e}}{\psi_1}$ in
$\varDelta_{i_1}$. The store instruction $\assign{i_1}{x}{e}$ may not
be simultaneously executed with $\assign{i_0}{r}{x}$. Therefore, for
any pre-condition $\varphi'_0$ that is sequentially composed on or
before the $\assign{i_0}{r}{x}$, we have to check that the effect of
$\assign{i_1}{x}{n}$ to the memory does not interfere $\varphi_0$ for
any $n \in \evp{e}{\varphi'_0 \wedge \varphi_1}$, which denotes
arbitrary values of $e$ that are constrained by $\varphi'_0 \wedge
\varphi_1$. For example, for $\assign{i_1}{x}{r_1}$ under $0 \leq r_1
< 3$, we have to consider the judgments about $\assign{i_1}{x}{0}$,
$\assign{i_1}{x}{1}$, and $\assign{i_1}{x}{2}$.

How do we check whether the effect of $\assign{i_1}{x}{n}$ to the
memory interferes $\varphi_0$ or not? A load instruction
$\assign{i_0}{r}{x}$ updates the vector clock of $i_0$, therefore,
$x^i$, and $r$. We divide it into two phases. If the updates of the
vector clock and $x^i$ are finished, the update of $r$ is sound. The
unsoundness of the assignment axiom for load instructions is derived
from updates of vector clocks.

Here, we do a trick.  By replacing thread identifier $i_1$ of the
store instruction with thread identifier $i_0$, we can update the
vector clock of thread identifier $i_0$. That is, for any $\varphi'_1$
that is a pre/post-condition on or after the $\assign{i_1}{x}{e}$, we
check $M \vDash^t \judge{\varphi'_1 \wedge
  \varphi_0}{\assign{i_0}{x}{n}}{\varphi_0}$ instead of any judgment
consisting of $\assign{i_1}{x}{n}$. Notably, our assertion language
cannot directly refer to vector clocks. We notice updates of vector
clocks through updates of observation variables.
Because judgments of store instructions are as-is sound, it is
sufficient to check $\vdash^t \judge{\varphi'_1 \wedge
  \varphi_0}{\assign{i_0}{x}{n}}{\varphi_0}$.

For example,
\[
\dfrac{
  {\judge{x^0 = 0}{\assign{0}{r}{x}}{r = 0}}\qquad
  {\judge{\top}{\assign{1}{x}{1}}{\top}}\qquad
  {\judge{\top \wedge x^0 = 0}{\assign{0}{x}{1}}{x^0 = 0}}
}{\judge{x^0 = 0}{\assign{0}{r}{x} \parallel \assign{1}{x}{1}}{r = 0}}
\]

\noindent
is not derivable because ${\judge{\top \wedge x^0 = 0}{\assign{0}{x}{1}}{x^0 = 0}}$ is not derivable.
Thus, we reuse the notion of non-interference for soundness of
judgments of load instructions.

\begin{table}[t]
  \caption{Remedying soundness.}\label{tab:soundness}%
{\footnotesize
  \begin{tabular}{l|c|c|c}
    \multirow{2}{*}{Without non-(observation-)interference} & \multicolumn{2}{c|}{assignment axiom} & \multirow{2}{*}{parallel composition rule}\\
    & \phantom{X}load\phantom{X} & store & \\\hline
    Owicki--Gries logic & \multicolumn{2}{c|}{$\checkmark$} & \\
    Our logic ($1$ thread) & $\checkmark$ & $\checkmark$ & N/A\\
    Our logic ($\geq 2$ threads) & & $\checkmark$ &\\
     & &\\
    \multirow{2}{*}{With non-(observation-)interference} & \multicolumn{2}{c|}{assignment axiom} & \multirow{2}{*}{parallel composition rule}\\
    & \phantom{X}load\phantom{X} & store & \\\hline
    Owicki--Gries logic & \multicolumn{2}{c|}{$\checkmark$} & $\checkmark$  \\
    Our logic ($1$ thread) & $\checkmark$ & $\checkmark$ & N/A\\
    Our logic ($\geq 2$ threads) & $\checkmark$ & $\checkmark$ & $\checkmark$
  \end{tabular}
}
\end{table}

The assignment axiom in Owicki--Gries logic is sound to the
standard state semantics, but the parallel composition rule without any
side condition is unsound. Therefore, Owicki and Gries introduced the
notion of non-interference.
The assignment axiom becomes unsound in our logic to timestamp
semantics if another thread executes a store
instruction. Non-observation-interference adopted by our logic plays a
role in keeping the assignment axiom for load instructions sound to
timestamp semantics.
Table~\ref{tab:soundness} summarizes it.

Next, we explain the extension of non-interference in our logic.
If an interfering instruction is $\assign{i_1}{r}{e}$, we have to
consider no issue in addition to non-interference. If an interfering
instruction is $\assign{i_1}{x}{e}$, we also have to consider no issue
because $x^{i_1}$ in the assertion possessed by thread $i_0$ is
synchronous and stability checking is the same as that in
Owicki--Gries logic, as explained at the beginning of
Section~\ref{sec:ours}.

If an interfering instruction is $\assign{i_1}{r}{x}$, we have to
take care of that $\assign{i_1}{r_1}{x}$ updates not only $r_1$ but
also $x^{i_1}$ in $\varphi_0$ and $\psi_0$.
Let $\judge{\varphi_0}{c_0}{\psi_0}$ be a possibly interfered judgment.
We have to confirm $M \vDash^t \judge{\varphi_1 \wedge
  \varphi_0}{\assign{i_1}{r_1}{x}}{\varphi_0}$ and $M \vDash^t
\judge{\varphi_1 \wedge \psi_0}{\assign{i_1}{r_1}{x}}{\psi_0}$.
Similar to the discussion about the soundness of judgments of load
instructions, we have to take care of that $\vdash^t \judge{\varphi_1
  \wedge \varphi_0}{\assign{i_1}{r_1}{x}}{\varphi_0}$ and $\vdash^t
\judge{\varphi_1 \wedge \psi_0}{\assign{i_1}{r_1}{x}}{\psi_0}$ are not
sufficient because they are not ensured not to be interfered with any
store instruction by another thread.

Thus, we have constructed logic that is sound to timestamp
semantics. However, the provability still needs to be
improved. Non-interference utilizes the pre-condition of the
interfering judgment, but it is impossible because the effect of the store
instruction may be delayed on timestamp semantics. It weakens the
expressive power of assertion language. Therefore, we cannot
delicately deal with execution traces; specifically, we cannot show the
correctness of Coherence under timestamp semantics because how to
grasp the difference between Store Buffering and Coherence, a total
order of timestamps at each variable on memory, has not been explained
yet.

We have introduced timestamp variables. Freshness in assignment axioms
of store instructions and non-logical axioms ensure total orders at
shared variables on memory. The variables are not updated by
assignments.  We can describe assertions. For any store instruction, a
fresh timestamp variable is defined and added to the end of the
sequence at each variable. It enables us to describe how far the
program runs. Timestamp variables are also used to restrict ranges
that instructions interfere in the definition of
non-observation-interference. Whereas auxiliary variables with
non-interference in Owicki--Gries logic do not restrict ranges that
instructions interfere with, pre-conditions that interfering
instructions have actually restricted ranges of interfered
conditions. In our logic, where we cannot assume pre-conditions for
interfering instructions, timestamp variables directly restrict ranges
of instructions that interfere with each other.

Finally, we formally define
non-observation-interference.  We call $\varDelta_i$s ($0 \leq i < N$)
non-observation-interfering if for any distinct $i_0$, $i_1$, and $i_2$,
\begin{itemize}
\item
  $\stable{t}{\vt{x}{t_0} < \vt{x}{t_1} \wedge
  \varphi'_1}{\assign{i_0}{x}{n}}{\varphi_0}$ holds for any
  $\judge{\varphi_0}{\assign{i_0}{r}{x}}{\_, \cons{\_}{\vt{x}{t_0}}}$
  in $\varDelta_{i_0}$, $\judge{\varphi_1}{\assign{i_1}{x}{e}}{\_,
    \cons{\_}{\vt{x}{t_1}}}$ in $\varDelta_{i_1}$, a pre-condition
  $\varphi'_0$ that is sequentially composed on or before the
  $\assign{i_0}{r}{x}$ in $\varDelta_{i_0}$, $n \in \evp{e}{\varphi'_0
    \wedge \varphi_1}$, and a pre/post-condition $\varphi'_1$ that is
  sequentially composed on or after the $\assign{i_1}{x}{e}$ in
  $\varDelta_{i_1}$,
\item $\stable{t}{\varphi_1}{\assign{i_1}{r}{e}}{\varphi_0}$ and
  $\stable{t}{\varphi_1}{\assign{i_1}{r}{e}}{\psi_0}$ hold for any
  $\judge{\varphi_0}{c_0}{\psi_0}$ in $\varDelta_{i_0}$ and
  $\judge{\varphi_1}{\assign{i_1}{r}{e}}{\_}$ in $\varDelta_{i_1}$,
\item $\stable{t}{\varphi_1}{\assign{i_1}{x}{e}}{\varphi_0}$ and
  $\stable{t}{\varphi_1}{\assign{i_1}{x}{e}}{\psi_0}$ hold for any
  $\judge{\varphi_0}{c_0}{\psi_0}$ in $\varDelta_{i_0}$ and
  $\judge{\varphi_1}{\assign{i_1}{x}{e}}{\_}$ in $\varDelta_{i_1}$, and
\item $\stable{t}{\varphi_1}{\assign{i_1}{r}{x}}{\varphi_0}$ and
  $\stable{t}{\varphi_1}{\assign{i_1}{r}{x}}{\psi_0}$ hold for any
  $\judge{\varphi_0}{c_0}{\psi_0}$ in $\varDelta_{i_0}$ and
  $\judge{\varphi_1}{\assign{i_1}{r}{x}}{\psi_1, \cons{\_}{\vt{x}{t_1}}}$ in
  $\varDelta_{i_1}$,
  moreover,
  \begin{itemize}
    \item $\stable{t}{\vt{x}{t_1} < \vt{x}{t_0} \wedge
      \varphi''_0}{\assign{i_1}{x}{n}}{\varphi_1 \wedge \varphi_0}$
      and $\stable{t}{\varphi''_0}{\assign{i_1}{x}{n}}{\varphi_1
        \wedge \psi_0}$ hold for any
      $\judge{\varphi'_0}{\assign{i_0}{x}{e}}{\_,
        \cons{\_}{\vt{x}{t_0}}}$ that is sequentially composed on or
      before the $c_0$ in $\varDelta_{i_0}$, a pre-condition
      $\varphi'_1$ that is sequentially composed on or before the
      $\assign{i_1}{r}{x}$ in $\varDelta_{i_1}$, a pre/post-condition
      $\varphi''_0$ that is sequentially composed on or after the
      $\assign{i_0}{x}{e}$ and on or before $c_0$ in
      $\varDelta_{i_0}$, and $n \in \evp{e}{\varphi'_1 \wedge
        \varphi'_0}$
    \item $\stable{t}{\vt{x}{t_1} < \vt{x}{t_2} \wedge
      \varphi'_2}{\assign{i_1}{x}{n}}{\varphi_1 \wedge \varphi_0}$ and
      $\stable{t}{\varphi'_2}{\assign{i_1}{x}{n}}{\varphi_1 \wedge
      \psi_0}$ hold for any $\judge{\varphi_2}{\assign{i_2}{x}{e}}{\_,
      \cons{\_}{\vt{x}{t_2}}}$ in $\varDelta_{i_2}$, a pre-condition
      $\varphi'_1$ that is sequentially composed on or before the
      $\assign{i_1}{r}{x}$ in $\varDelta_{i_1}$, $n \in
      \evp{e}{\varphi'_1 \wedge \varphi_2}$, and a pre/post-condition
      $\varphi'_2$ that is sequentially composed on or after the
      $\assign{i_2}{x}{e}$ in $\varDelta_{i_2}$
  \end{itemize}
\end{itemize}
where we write
$\stable{t}{\varphi_1}{c}{\varphi_0}$ for $\vdash^t \judge{\varphi_1
  \wedge \varphi_0, \_}{c}{\varphi_0, \_}$, for short.

\begin{theorem}
  If $\rho$ follows $\tvl_{N-1}$, then $\vdash^t \judge{\varphi}{s_0
    \parallel \cdots \parallel s_{N-1}}{\psi, \tvl_{0},\ldots
    \tvl_{N-1}}$ implies $\pair{\{\pair{\_}{0} \mapsto
    0\}}{\pair{\mathfrak{S}}{\{\_ \mapsto \{\_ \mapsto 0\}\}}}, \rho
  \vDash^t \judge{\varphi}{s_0 \parallel \cdots \parallel
    s_{N-1}}{\psi, \tvl_0, \ldots, \tvl_{N-1}}$ for any
  $\mathfrak{S}$.
\end{theorem}

{
\newcommand{\oneone}{r_0 = 0 \wedge r_1 \leq x^1}
\newcommand{\onetwo}{0 < x^0 \wedge r_0 = 0 \wedge r_1 \leq x^1}
\newcommand{\onethree}{0 < r_0 }
\newcommand{\onecomtwo}{\assign{0}{r_0}{x}}
\newcommand{\onecomone}{\assign{0}{x}{1}}
\newcommand{\ponecomone}{\assign{1}{x}{1}}
\newcommand{\twoone}{r_1 = 0 \wedge r_0 \leq x^0 \wedge x^0 \leq \vt{x}{t_1} }
\newcommand{\twotwo}{r_1 = 0 \wedge r_0 \leq x^0 \wedge x^0 \leq \vt{x}{t_1} \wedge \vt{x}{t_1} \leq x^1}
\newcommand{\twothree}{r_0 \leq x^0 \wedge x^0 \leq \vt{x}{t_1} \wedge \vt{x}{t_1} \leq r_1}
\newcommand{\twocomone}{\assign{1}{x}{2}}
\newcommand{\ptwocomone}{\assign{0}{x}{2}}
\newcommand{\twocomtwo}{\assign{1}{r_1}{x}}

\begin{figure}
{\small
\lstset{frame=none}
\begin{minipage}{.14\textwidth}
\begin{lstlisting}
 x := 1
 r0 := x
\end{lstlisting}
\end{minipage}
%\;\makebox[0pt][c]{\raisebox{15pt}{$\xymatrix@C=160pt{\ar@{-}[rd]&\ar@{-}[ld]\\&}$}}
\lstset{frame=L}
\begin{minipage}{.10\textwidth}
\begin{lstlisting}
 x := 2
 r1 := x
\end{lstlisting}
\end{minipage}
}
{\footnotesize
\begin{minipage}{.28\textwidth}
  \[
  \begin{array}{c}
    \smash{\oneone}\rule{0pt}{2.5ex}\\
    \smash{\{\onecomone\}}\rule{0pt}{2.5ex}\\
    \smash{\onetwo}\rule{0pt}{2.5ex}\\
    \smash{\{\onecomtwo\}}\rule{0pt}{2.5ex}\\
    \smash{\onethree}\rule{0pt}{2.5ex}
  \end{array}
  \]
\end{minipage}
\begin{minipage}{.24\textwidth}
  \[
  \begin{array}{c}
    \smash{\twoone}\rule{0pt}{2.5ex}\\
    \smash{\{\twocomone\}}\rule{0pt}{2.5ex}\\
    \smash{\twotwo}\rule{0pt}{2.5ex}\\
    \smash{\{\twocomtwo\}}\rule{0pt}{2.5ex}\\
    \smash{\twothree}\rule{0pt}{2.5ex}
  \end{array}
  \]
\end{minipage}

\vspace{\baselineskip}
invariant: $(x^0 = 0 \vee x^0 = 1 \vee x^0 = 2) \wedge (x^1 = 0 \vee x^1 = 1 \vee x^1 = 2) \wedge (r_0 = 0 \vee r_0 = 1 \vee r_0 = 2) \wedge (r_1 = 0 \vee r_1 = 1 \vee r_1 = 2) \wedge (\vt{x}{t_0} = 1 \vee \vt{x}{t_0} = 2) \wedge (\vt{x}{t_1} = 1 \vee \vt{x}{t_1} = 2) \wedge (\vt{x}{t_0} < \vt{x}{t_1} \vee \vt{x}{t_1} < \vt{x}{t_0})$ where $\vt{x}{t_0}$ and $\vt{x}{t_1}$ are freshly generated at $\assign{0}{x}{1}$ and $\assign{1}{x}{2}$ in the judgments, respectively.
}
\caption{Derivation of Coherence on non-observation-interference.}\label{fig:ch_OB}
\end{figure}

\begin{figure}
{\scriptsize
  $\stable{t}{\vt{x}{t_0} < \vt{x}{t_1} \wedge \twotwo}{\ptwocomone}{\oneone}$

  $\stable{t}{\vt{x}{t_0} < \vt{x}{t_1} \wedge \twothree}{\ptwocomone}{\oneone}$

  $\stable{t}{\twoone}{\twocomone}{\oneone}$

  $\stable{t}{\twoone}{\twocomone}{\onetwo}$

  $\stable{t}{\twoone}{\twocomone}{\onethree}$

  $\stable{t}{\twotwo}{\twocomtwo}{\oneone}$

  $\stable{t}{\twotwo}{\twocomtwo}{\onetwo}$

  $\stable{t}{\twotwo}{\twocomtwo}{\onethree}$

  $\stable{t}{\vt{x}{t_1} < \vt{x}{t_0} \wedge \oneone}{\ponecomone}{\twotwo \wedge \oneone}$

  $\stable{t}{\vt{x}{t_1} < \vt{x}{t_0} \wedge \oneone}{\ponecomone}{\twotwo \wedge \onetwo}$

  $\stable{t}{\vt{x}{t_1} < \vt{x}{t_0} \wedge \oneone}{\ponecomone}{\twotwo \wedge \onethree}$

  $\stable{t}{\vt{x}{t_1} < \vt{x}{t_0} \wedge \onetwo}{\ponecomone}{\twotwo \wedge \oneone}$

  $\stable{t}{\vt{x}{t_1} < \vt{x}{t_0} \wedge \onetwo}{\ponecomone}{\twotwo \wedge \onetwo}$

  $\stable{t}{\vt{x}{t_1} < \vt{x}{t_0} \wedge \onetwo}{\ponecomone}{\twotwo \wedge \onethree}$

  $\stable{t}{\vt{x}{t_1} < \vt{x}{t_0} \wedge \onethree}{\ponecomone}{\twotwo \wedge \oneone}$

  $\stable{t}{\vt{x}{t_1} < \vt{x}{t_0} \wedge \onethree}{\ponecomone}{\twotwo \wedge \onetwo}$

  $\stable{t}{\vt{x}{t_1} < \vt{x}{t_0} \wedge \onethree}{\ponecomone}{\twotwo \wedge \onethree}$
}
\vspace{\baselineskip}

{\scriptsize
  $\stable{t}{\vt{x}{t_1} < \vt{x}{t_0} \wedge \onetwo}{\ponecomone}{\twoone}$

  $\stable{t}{\vt{x}{t_1} < \vt{x}{t_0} \wedge \onethree}{\ponecomone}{\twoone}$

  $\stable{t}{\oneone}{\onecomone}{\twoone}$

  $\stable{t}{\oneone}{\onecomone}{\twotwo}$

  $\stable{t}{\oneone}{\onecomone}{\twothree}$

  $\stable{t}{\onetwo}{\onecomtwo}{\twoone}$

  $\stable{t}{\onetwo}{\onecomtwo}{\twotwo}$

  $\stable{t}{\onetwo}{\onecomtwo}{\twothree}$

  $\stable{t}{\vt{x}{t_0} < \vt{x}{t_1} \wedge \twoone}{\ptwocomone}{\onetwo \wedge \twoone}$

  $\stable{t}{\vt{x}{t_0} < \vt{x}{t_1} \wedge \twoone}{\ptwocomone}{\onetwo \wedge \twotwo}$

  $\stable{t}{\vt{x}{t_0} < \vt{x}{t_1} \wedge \twoone}{\ptwocomone}{\onetwo \wedge \twothree}$

  $\stable{t}{\vt{x}{t_0} < \vt{x}{t_1} \wedge \twotwo}{\ptwocomone}{\onetwo \wedge \twoone}$

  $\stable{t}{\vt{x}{t_0} < \vt{x}{t_1} \wedge \twotwo}{\ptwocomone}{\onetwo \wedge \twotwo}$

  $\stable{t}{\vt{x}{t_0} < \vt{x}{t_1} \wedge \twotwo}{\ptwocomone}{\onetwo \wedge \twothree}$

  $\stable{t}{\vt{x}{t_0} < \vt{x}{t_1} \wedge \twothree}{\ptwocomone}{\onetwo \wedge \twoone}$

  $\stable{t}{\vt{x}{t_0} < \vt{x}{t_1} \wedge \twothree}{\ptwocomone}{\onetwo \wedge \twotwo}$

  $\stable{t}{\vt{x}{t_0} < \vt{x}{t_1} \wedge \twothree}{\ptwocomone}{\onetwo \wedge \twothree}$
}
\caption{Stability checking of Coherence on non-observation-interference.}\label{fig:ch_OB2}
\end{figure}
}

Figures~\ref{fig:ch_OB} and \ref{fig:ch_OB2} show a derivation of Coherence on
non-observation-interference.
We can see that the timestamp variable
$\vt{x}{t_1}$ acts as an intermediary between $r_0$ and $r_1$.

\subsection{Auxiliary Variable Extension}\label{sec:aux}

In order to consider a delay in the effect of $\assign{i}{x}{e}$, we
take any $n \in \evp{e}{\varphi}$ for some $\varphi$. Because
pre-conditions are not invariant, non-inteference checking cannot
continue to assume $\varphi$. The unconnectiveness of values and
pre-conditions weakens the provability. For example, although $r_0 =
0$ holds after executing the following program

\lstset{frame=none}
{\centering
\small
\begin{minipage}{75pt}
\begin{lstlisting}
 r0 := x  // 1
 y := 1
\end{lstlisting}
\end{minipage}
\quad
\lstset{frame=L}
\begin{minipage}{70pt}
\begin{lstlisting}
 r1 := y
 x := r1
\end{lstlisting}
\end{minipage}\par
}

\noindent
on timestamp semantics in this paper, we have to consider both of the
effects $\assign{0}{x}{0}$ and $\assign{0}{x}{1}$ raised by the
execution of $\assign{1}{x}{r_1}$ without any hints.

As explained in Section~\ref{sec:og}, auxiliary variables enhance the
provability of concurrent program logic. Auxiliary variables are
written only once and are write-only in this paper. Any assignment to
an auxiliary variable is atomically executed with an assignment
statement, that is, any interleaving by an instruction. Auxiliary
variables are prohibited to be read in programs in order not to change
behaviors of programs. We use auxiliary variables for describing
assertions.

We modify the first and fourth conditions of the non-interference
checking as follows:
\begin{itemize}
\item
  $\stable{t}{\vt{x}{t_0} < \vt{x}{t_1} \wedge \varphi'_1 \wedge
  \varphi}{\assign{i_0}{x}{n}, \assign{i_1}{a_0}{n_0}}{\varphi_0}$
  holds for any $\judge{\varphi_0}{\assign{i_0}{r}{x}}{\_,
    \cons{\_}{\vt{x}{t_0}}}$ in $\varDelta_{i_0}$,
  $\judge{\varphi_1}{\assign{i_1}{x}{e}, \assign{i_1}{a_0}{n_0}}{\_,
    \cons{\_}{\vt{x}{t_1}}}$ in $\varDelta_{i_1}$, a pre-condition
  $\varphi'_0$ that is sequentially composed on or before the
  $\assign{i_0}{r}{x}$ in $\varDelta_{i_0}$, $n \in \evp{e}{\varphi'_0
    \wedge \varphi_1}$, a pre/post-condition $\varphi'_1$ that is
  sequentially composed on or after the $\assign{i_1}{x}{e}$ in
  $\varDelta_{i_1}$, and $\varphi$ is any onjunction of
  \begin{itemize}
  \item the equations $\psi$ between the
  auxiliary variables and their values when the interpretation of $e$
  by $\varphi'_0 \wedge \varphi_1$ is $n$, or
  \item the equations between the auxiliary variables and their
    specified values that are set to the auxiliary variables more than
    $\psi$,
  \end{itemize}
\item $\stable{t}{\varphi_1}{\assign{i_1}{r}{x}}{\varphi_0}$ and
  $\stable{t}{\varphi_1}{\assign{i_1}{r}{x}}{\psi_0}$ hold for any
  $\judge{\varphi_0}{c_0}{\psi_0}$ in $\varDelta_{i_0}$ and
  $\judge{\varphi_1}{\assign{i_1}{r}{x}}{\psi_1,
  \cons{\_}{\vt{x}{t_1}}}$ in $\varDelta_{i_1}$, moreover,
  \begin{itemize}
    \item $\stable{t}{\vt{x}{t_1} < \vt{x}{t_0} \wedge \varphi''_0
      \wedge \varphi}{\assign{i_1}{x}{n},
      \assign{i_0}{a_0}{n_0}}{\varphi_1 \wedge \varphi_0}$ and
      $\stable{t}{\varphi''_0 \wedge \varphi}{\assign{i_1}{x}{n},
      \assign{i_0}{a_0}{n_0}}{\varphi_1 \wedge \psi_0}$ hold for any
      $\judge{\varphi'_0}{\assign{i_0}{x}{e},
      \assign{i_0}{a_0}{n_0}}{\_, \cons{\_}{\vt{x}{t_0}}}$ that is
      sequentially composed on or before the $c_0$ in
      $\varDelta_{i_0}$, a pre-condition $\varphi'_1$ that is
      sequentially composed on or before the $\assign{i_1}{r}{x}$ in
      $\varDelta_{i_1}$, a pre/post-condition $\varphi''_0$ that is
      sequentially composed on or after the $\assign{i_0}{x}{e}$ and
      on or before $c_0$ in $\varDelta_{i_0}$, $n \in
      \evp{e}{\varphi'_1 \wedge \varphi'_0}$, and $\varphi$ is any
      conjunction of
      \begin{itemize}
      \item the equations $\psi$ between the auxiliary variables and their
        values when the interpretation of $e$ by $\varphi'_1 \wedge
        \varphi'_0$ is $n$, or
      \item the equation between the auxiliary variables and their
        specified values that are set to the auxiliary variables more
        than $\psi$,
      \end{itemize}
    \item $\stable{t}{\vt{x}{t_1} < \vt{x}{t_2} \wedge \varphi'_2
      \wedge \varphi}{\assign{i_1}{x}{n},
      \assign{i_2}{a_0}{n_0}}{\varphi_1 \wedge \varphi_0}$ and
      $\stable{t}{\varphi'_2 \wedge \varphi}{\assign{i_1}{x}{n},
      \assign{i_2}{a_0}{n_0}}{\varphi_1 \wedge \psi_0}$ hold for any
      $\judge{\varphi_2}{\assign{i_2}{x}{e},
      \assign{i_2}{a_0}{n_0}}{\_, \cons{\_}{\vt{x}{t_2}}}$ in
      $\varDelta_{i_2}$, a pre-condition $\varphi'_1$ that is
      sequentially composed on or before the $\assign{i_1}{r}{x}$ in
      $\varDelta_{i_1}$, $n \in \evp{e}{\varphi'_1 \wedge \varphi_2}$,
      a pre/post-condition $\varphi'_2$ that is sequentially composed
      on or after the $\assign{i_2}{x}{e}$ in $\varDelta_{i_2}$, and
      $\varphi$ is any conjunction of
      \begin{itemize}
      \item the equations $\psi$ between the auxiliary variables and their
        values when the interpretation of $e$ by $\varphi'_1 \wedge
        \varphi_2$ is $n$, or
      \item the equations between the auxiliary variables and their
        specified values that are set to the auxiliary variables more
        than $\psi$.
      \end{itemize}
  \end{itemize}
\end{itemize}

\noindent
It is noteworthy that we assume that the specified values are again
set to the auxiliary variables in loading the delayed effects. It
helps to enhance the provability of our logic.
We take advantage of the fact that auxiliary variables are never
reset to their initial values.

%Although the value of an auxiliary variable is set when the
%assignment to the auxiliary variable is executed,

\begin{figure}
{
\newcommand{\oneone}{x^0 = 0 \wedge a_0 = 0}
\newcommand{\onetwo}{r_0 = x^0 \wedge x^0 = 0}
\newcommand{\onethree}{r_0 = 0}
\newcommand{\onecomone}{\assign{0}{r_0}{x}}
\newcommand{\onecomtwo}{\assign{0}{y}{1}, \assign{0}{a_0}{1}}
\newcommand{\twoone}{y^1 \leq a_0}
\newcommand{\twotwo}{r_1 \leq a_0}
\newcommand{\twothree}{\top}
\newcommand{\twocomone}{\assign{1}{r_1}{y}}
\newcommand{\twocomtwo}{\assign{1}{x}{r_1}}
{
\centering
\lstset{frame=none}
%{\centering
\small
\begin{minipage}{50pt}
\begin{lstlisting}
 r0 := x
 atomic {
  y := 1
  a0 := 1
 }
\end{lstlisting}
\end{minipage}
\quad
\lstset{frame=L}
\begin{minipage}{60pt}
\begin{lstlisting}
 r1 := y
 x := r1
#\phantom{X}#
#\phantom{X}#
#\phantom{X}#
\end{lstlisting}
\end{minipage}
%}
\footnotesize
\begin{minipage}{.18\textwidth}
  \[
  \begin{array}{c}
    \smash{\oneone}\rule{0pt}{2.5ex}\\
    \smash{\{\onecomone\}}\rule{0pt}{2.5ex}\\
    \smash{\onetwo}\rule{0pt}{2.5ex}\\
    \smash{\{\onecomtwo\}}\rule{0pt}{2.5ex}\\
    \smash{\onethree}\rule{0pt}{2.5ex}
  \end{array}
  \]
\end{minipage}
\qquad
\begin{minipage}{.18\textwidth}
  \[
  \begin{array}{c}
    \smash{\twoone}\rule{0pt}{2.5ex}\\
    \smash{\{\twocomone\}}\rule{0pt}{2.5ex}\\
    \smash{\twotwo}\rule{0pt}{2.5ex}\\
    \smash{\{\twocomtwo\}}\rule{0pt}{2.5ex}\\
    \smash{\twothree}\rule{0pt}{2.5ex}
  \end{array}
  \]
\end{minipage}

\noindent
where the global invariant $(x^0 = 0 \vee x^0 = 1) \wedge (x^1 = 0 \vee x^1 = 1) \wedge (y^0 = 0 \vee y^0 = 1) \wedge (y^1 = 0 \vee y^1 = 1) \wedge (r_0 = 0 \vee r_0 = 1) \wedge (r_1 = 0 \vee r_1 = 1)$ is omitted.
}
}
\caption{Derivation using an auxiliary variable.}\label{fig:klb}
\end{figure}

We attach $\assign{0}{a_0}{1}$ with $\assign{0}{y}{1}$ in the above
example program where $a_0$ is an auxiliary variable.
Figure~\ref{fig:klb} is an example derivation using an auxiliary
variable where the non-interference checking is omitted.
By taking the pre-condition $r_1 \leq a_0$ of the assignment statement
$\assign{1}{x}{r_1}$, it is unnecessary to consider the effect
$\assign{0}{x}{1}$ of the store instruction $\assign{1}{x}{r_1}$ under
$a_0 = 0$, although it is surely necessary to consider the effect
$\assign{0}{x}{1}$ of the store instruction $\assign{1}{x}{r_1}$ under
$a_0 = 1$.

\section{Related Work and Discussion}\label{sec:related}

Some logics are sound to semantics on which
the effects of store instructions may be
delayed~\cite{DBLP:conf/vstte/Ridge10,conf/icalp/LahavV15,am2016aplas,DDDW20}.

Dalvandi et al.\ provided Owicki--Gries logic for a C11 fragment
memory model~\cite{DDDW20}. They also adopted timestamp
semantics. They introduced multiple relations based on observations by
thread to their assertion language. Definite observation $x
\mathrel{=_i} n$ denotes that thread $i$ must observe value $n$ for
$x$. The idea is different from our idea that assertions are described
by observation variables and the standard equality. Possible
observation $x \approx_i n$ denotes that thread $i$ may observe value
$n$ for $x$. Conditional observation $[x = m] (y \mathrel{=_i} n)$
denotes that if $x=m$ is synchronized, thread $i$ must observe value
$n$ for $y$. These relations are used to support the
release-acquire synchronization on the C11 memory model. They also
introduced definite and possible value orders, which denote orders
between the effects of store instructions like our timestamp
variables. We also took an approach of describing formulas in such a
rich assertion language and giving them to a model
checker~\cite{am2017sttt}. However, in this study, we aim to construct
theoretically simple logic in the hope that our logic would be used by
a basis for concurrent program logic for various memory models, and
adopted the idea that assertions are described by observation
variables, timestamp variables, and the standard equality.

Lahav and Vafeiadis proposed Owicki--Gries style logic for semantics
by execution graphs~\cite{conf/icalp/LahavV15}, more axiomatic
semantics than operational timestamp semantics. Their logic has high
provability. They succeeded in providing derivations to ensure the
correctness of Store Buffering with memory fences and a variant of
Coherence. They also modified stability checking to ensure the
correctness of concurrent programs on a weak memory model, as we did
in this paper. Their logic differs from ours because they are
constructed without observation and timestamp variables.

An unsatisfactory point is that it is necessary to check judgments of
the form $\varGamma \not\vdash \varphi$ in their logic. Specifically,
the logic requires finding satisfiable values under the semantics when
checking stability.  This means that it is necessary to consider
models when providing derivations.
Dalvandi et~al.\ adopts assertions that describe global behaviors
of programs; for example, $\mathbbm{1}_x 2$ denotes that there exists
at most one write with $2$ to $x$~\cite{DDDW20}.
Whereas the author provided logic using observation
variables, the logic also assumes the use of
external knowledge called observation invariants~\cite{am2016aplas}.

Thus, semantics on which effects of store instructions are delayed
often disturb us to construct purely syntactical logic in which
assertions are added to statements as pre/post-conditions in the Floyd
style~\cite{floyd1967}. In our logic, it occurs as the arbitrariness
of the values of store instructions in the definition of
non-observation-interference as described in
Section~\ref{sec:ours}. This study is one of the challenges to finding
purely syntactical logic for weak memory models.

The author proposed observation variables in the previous
work~\cite{am2016aplas}. However, the observation variables of a
shared variable do not almost interact with one another in the logic.
Observation invariants, which are assumed to be externally given,
specify relations between the observation variables of a shared
variable. In this paper, such relations are internally described in
the definition of non-observation-interference. This work is
positioned as a detailed investigation of shared variables observed by
threads.

\section{Conclusion and Future Work}\label{sec:conclusion}

In this paper, we provide Owicki--Gries style logic, which uses the
notion of non-interference for timestamp semantics. It theoretically
contributes to showing the techniques that we 1)used the notion of
non-interference for ensuring soundness of the assignment axiom for
load instructions, 2)replaced thread identifiers for an update of a
vector clock on timestamp semantics, and 3)introduced timestamp
variables in order to enhance the expressiveness of the assertion
language.

This study has a lot left to do. In this paper, we do not refer to
completeness to timestamp semantics.
We do not confirm the minimality of assertions that occur in the
derivations in this paper.
Above all else, Owicki--Gries logic using non-interference is not
compositional; the depths of derivations combinatorially increase the
number of checking judgments. It means that Owicki--Gries logic and
our logic could be more practical. Jones proposed a compositional
logic using rely/guarantee reasoning~\cite{Jones1981}. Lahav and
Vafeiadis tried to make their logic compositional using the
rely/guarantee notion. However, because their ``rely'' and
``guarantee'' are not single assertions but finite sets of
assertions~\cite{conf/icalp/LahavV15}, it is hard to say that
combinatorial explosion is wholly avoided.

Coughlin et al.\ suggested the weak point and constructed a
rely/guarantee system for multicopy atomic weak memory
models~\cite{Coughlin2021}. They regard the Owicki--Gries interference
as horizontal interference between threads, and proposed reordering
interference as vertical interference within one thread. Let $R, G
\vdash \judge{\varphi}{c}{\chi}$ and $R, G \vdash
\judge{\chi}{c'}{\psi}$ be derivable judgments in their logic. If $c;
c'$ may be reordered to $c'_{\langle c \rangle}; c$, their reordering
interference freedom checking requires $\chi'$ such that $R, G \vdash
\judge{\varphi}{c'_{\langle c \rangle}}{\chi'}$ and $R, G \vdash
\judge{\chi'}{c}{\psi}$ are derivable.

In our case to adopt not reordering of instructions but observation
variables, enriching the assertion language using the next variables
(e.g.~\cite{DBLP:journals/fac/XuRH97}) is promising. Our previous
study also proposed a compositional logic using observation
variables~\cite{am2016aplas}, so to speak, observation-based Jones
logic.  However, external knowledge is required, called observation
invariants, as described in Section~\ref{sec:related}. Thus, it is
challenging to construct simple, reasonable, compositional logic for
weak memory models.  We try to construct a logic using observation
variables by utilizing the knowledge obtained in this study.

This study considered store buffering only. It is not easy to consider
load buffering. Whereas we previously proposed concurrent program
logic sound and complete to semantics that support reorderings of load
instructions~\cite{am2017jip}, the semantics support only reorderings
of load instructions that are statically determined to be
independent. For a weaker memory model, Kang et al.\ proposed a novel
notion called promise for supporting load buffering on timestamp
semantics~\cite{conf/popl/Kang2017}. For example, on timestamp
semantics in this paper, assertion $r_0 = 0$ holds after the program
in Section~\ref{sec:aux} is executed.  However, $r_0 = 1$ is allowed
on timestamp semantics with promise. We try to combine the notion of
promise with our logic and construct a novel logic to ensure the
correctness of concurrent programs on semantics that allow load
buffering.

\vspace{\baselineskip}
\noindent
\textbf{Acknowledgments.}
This work was supported by JSPS KAKENHI Grant Number JP23K11051.

\bibliographystyle{abbrv}
\bibliography{draft}

\end{document}